\documentclass{article}

\usepackage{amsmath}
\usepackage{amssymb}
\usepackage{rotfloat}
\usepackage{dsfont}
\usepackage{hyperref}
\usepackage[usenames,dvipsnames]{color}
\usepackage{multirow}
\usepackage{graphicx}

\begin{document}

\newcommand{\SBE}{RE}
\newcommand{\PE}{PE}
\newcommand{\CE}{CE}
\newcommand{\ME}{MME}
\newcommand{\BEV}{BEV}
\newcommand{\BEC}{BEC}
\newcommand \myin     {\mbox{\,$\in$}\,}
\newcommand \Forall   {{\forall}\,}
\newcommand \Exists   {{\exists}\,}
\newcommand \NAT      {\mathds{N}}
\newcommand \INT      {\mathds{Z}}
\newcommand \RAT      {\mathds{Q}}
\newcommand \REAL     {\mathds{R}}
\newcommand \COMPL    {\mathds{C}}
\newcommand \FS       {\mathfrak{S}}
\newcommand \Rank     {\textrm{rank}}

\newcommand{\remark}[1]{\textcolor{red}{#1}}


\title{Solving Hard Control Problems in Voting Systems \\ via Integer Programming}

\author{Sergey Polyakovskiy\\
Optimisation and Logistics\\
School of Computer Science \\
             The University of Adelaide \\
             Adelaide, SA 5005, Australia.
\and
Rudolf Berghammer\\
Institut f\"ur Informatik \\
             Christian-Albrechts-Universit\"at zu Kiel \\ 
             Olshausenstra\ss{}e 40, 24098 Kiel, Germany.
\and
Frank Neumann\\
Optimisation and Logistics\\
School of Computer Science \\
             The University of Adelaide \\
             Adelaide, SA 5005, Australia.
}

\maketitle

\begin{abstract}
Voting problems are central in the area of social choice. 
In this article, we investigate various voting systems and types of 
control of elections. We present integer 
linear programming (ILP) formulations for a wide range of NP-hard 
control problems. Our ILP formulations are 
flexible in the sense that  they can work with an arbitrary number 
of candidates and voters.
Using the off-the-shelf solver \textsc{Cplex}, we show that our approaches
can manipulate elections with a large number of voters and candidates efficiently.
\end{abstract}

\section{Introduction}\label{Sec1}

When a group of people with individual preferences has to decide which 
alternative to choose from a given set of alternatives, an election is 
often carried out.
The voting rule underlying the election can be regarded 
as an algorithm that computes from the individual preferences of the people
(which in this context are called voters) those alternatives which are 
accepted as `best' choices by the whole group.
Ideally, there should be exactly one such alternative, the winning one.

There are many different voting rules to determine the winners 
of elections, each coming with different advantages and drawbacks. 
For example, it is desirable to have a voting rule that can be 
computed efficiently and has exactly one winner.
On the other hand, if it is easy to manipulate the election structure to 
get one's favorite candidate to win, 
then this is regarded as negative in view of susceptibility to 
illegal influence. \cite{Walsh} discusses the types of \emph{illegal influence}.
So, voting rules not only should be efficient, they also should be hard 
(ideally even impossible) to influence in an illegal way.

Research on computational social choice applies techniques of computer science, mainly 
from algorithmic and complexity theory, to problems from social choice
theory (see the articles of \cite{Chevaleyre} and \cite{BraConEnd} for an overview). 
Central computational questions in the area of computational social choice are
the efficiency of voting rules and their susceptibility to illegal influence.
In this paper, we investigate
a specific kind of illegal influence, called \emph{control}.
We investigate the case where an actor seeks to have a desired 
candidate winning the election by removing a set of voters 
(or candidates) from the election.

\cite{BarTovTri} have initiated a new line of research that investigates
the susceptibility to control by techniques from complexity theory.
The goal is to prevent attacks using certain types of control by showing
that they lead to NP-hard decision problems.
Following \cite{BarTovTri}, numerous papers have investigated the complexity 
of control problems for elections 
(\cite{ConitzerSandholgLang}, \cite{HemHemRot}, 
\cite{FaHeSc}, \cite{FalHemHem}, and \cite{ErdFelRotSch}).
For many voting systems it is shown that certain control problems are hard.
From a theoretical point of view such voting systems can be regarded as 
secure against this attempt to illegal influence.
The hardness results mentioned previously often assume a growing number of candidates and voters.
It is known that many voting problems with few candidates are easy to manipulate~(\cite{ConitzerSandholgLang,DBLP:conf/aaai/Walsh07}) as this restricts the number of choices the voters have.

In the context of computational social choice, \cite{Walsh} clearly demonstrates
that NP-hardness is not a barrier to manipulations and 
illegal influences. \cite{ConitzerSandholm} present a simple 
influence algorithm for elections
 that works fast and yields for most inputs (according to a suitably 
chosen probability distribution) the desired result.
Another approach to solve hard control problems in practice is proposed
by \cite{BerDanSch}, \cite{BerSch}, and \cite{Berghammer}.
It combines relation algebra and the BDD-based computer algebra system 
\textsc{RelView} and yields algorithms which are correct for all instances.
Different problems from social choice have also been examined 
in the context of fixed-parameter tractability, and fixed parameter 
algorithms can be obtained in many cases by using integer 
programming formulations 
(see for example \cite{Lindner08fixed-parametertractability,Hemaspaandra20131345}).
For rules like Borda, Bucklin, and Copeland, it is fixed-parameter 
tractable with respect to the number of candidates to determine 
possible winners when given are incomplete votes 
(\cite{DBLP:conf/ijcai/BetzlerHN09}). 

We study voting problems in terms of the number of voters $n$ 
and the number of candidates $m$, where both are variable.
The present paper follows the line initiated by \cite{GurRoo} who used 
binary integer programming to solve some hard
control problems for two closely related voting systems, known as 
\emph{Copeland voting} and \emph{Llull voting}, respectively. 
We show how the hard control 
problems of the (quite different) voting systems can 
be specified as integer programs, and present results of computational 
experiments. Our computational results show a good performances of the 
introduced approaches even if both the number of voters and the 
number of candidates is large.

The remainder of the paper is organized as follows. In Section~\ref{Sec2}, 
we introduce the notion of a voting system and present
the specific voting systems we will consider in this paper.
Section~\ref{Sec3} is devoted to several control problems in 
voting systems and their computational complexities. How to model the hard 
types of control for the voting systems of Section~\ref{Sec2}
as integer programs is shown in Section~\ref{Sec4}, which constitutes the core 
of the paper. Herein, we mainly concentrate on constructive control where the actor 
seeks to ensure his favourite alternative's victory.
Section~\ref{Sec5} presents the results of our computational experiments, 
which we have performed on the test benchmark suite of the Preference Library 
(commonly referred to as PrefLib, \cite{PrefLib}) using one of the well-known
off-the-shelf solvers for integer programs. 
The results of our experiments show that our approach is able to solve all 
the test instances to optimality quite fast using an ordinary personal computer. 
Moreover, even the hard instances related to larger elections can be handled 
in reasonable time. 
Finally, Section~\ref{Sec6} contains some concluding remarks and presents 
topics for future investigations in this area of work.

\section{Voting Systems}\label{Sec2}

In the following, we introduce the notions of a voting system and an election 
as generally used in social choice theory as well as the particular voting 
systems we will treat in this paper. 
For more details on voting in social choice theory and additional voting 
systems we refer to the studies of \cite{Tideman}, \cite{BraFis}, \cite{Las}, 
and \cite{BraConEnd}.

In social choice theory, a \emph{voting system} (also called a \emph{voting 
protocol}) consists of a finite and non-empty set $C$ of \emph{candidates} 
(alternatives, proposals, options), a finite and non-empty set $V$ of 
\emph{voters} (players, agents, individuals), the \emph{individual 
preferences} (choices, wishes) of the single voters, and a \emph{voting 
rule} that aggregates the winners from the individual preferences. 
Usually the pair $(C,V)$ is called an \emph{election}.
In such a definition the representation of the individual preferences remains
out of consideration.
Since it, however, will play a fundamental role in our approach, in this
paper we use a slightly more general notion of elections and define them as
triples $(C,V,I)$, where the third component $I$ is the specification of 
the individual preferences.

To simplify presentation, in the remainder of the paper we assume that the two 
sets $C$ and $V$ are given as $C =\{c_1,\ldots,c_m\}$ and 
$V = \{v_1,\ldots,v_n\}$, where $m, n \myin \NAT_{>0}$ are non-zero natural 
numbers.

\subsection{Approval Voting}\label{SubSec21}

A well-known voting system is \emph{approval voting}, which has been introduced by \cite{BraFis}. 
For example, it is presently used by several scientific organizations including the
Mathematical Association of America and the Institute of Management Science. 
Here each voter may approve (that is, vote for) as many candidates as he wants 
and then the candidate with more approvals than all other candidates is declared 
as the winner. 
If we model the individual preferences by functions $a_v : C \to \{0,1\}$ such 
that $a_v(c) = 1$ iff voter $v$ approves candidate $c$, for all $v \myin V$ and 
$c \myin C$, then an approval election can be specified formally as a 
triple $(C,V,(a_v)_{v \myin V})$ and a candidate $c^*$ is then defined as the 
winner iff 
\begin{equation}
\sum_{v \myin V} a_v(c^*) > \sum_{v \myin V} a_v(c),
\label{WINAPP}
\end{equation}
for all $c \myin C \setminus \{c^*\}$.
Note, that this specification of a winner implies winners to be unique. 
In case that inequality (\ref{WINAPP}) holds, candidate $c^*$ is said 
\emph{to strictly dominate} candidate $c$. 

The strict dominance relation $D$ on the set $C$, for all $c,d \in C$ defined by
$c\, D \, d$ iff $\sum_{v \myin V} a_v(c) > \sum_{v \myin V} a_v(d)$, is asymmetric.
But it may happen that there exist pairs of different candidates $c$ and $d$
such that neither $c\, D \, d$ nor $d\, D \, c$.
As a consequence, a candidate that strictly dominates all other ones (the winner) 
does not necessarily exist. 
For this reason, in the literature a variant of approval voting is also investigated, 
where dominance is weak. 
Here, candidate $c^*$ wins the election $(C,V,(a_v)_{v \myin V})$ iff for all 
$c \myin C$ it holds $\sum_{v \myin V} a_v(c^*) \geq \sum_{v \myin V} a_v(c)$. 
The advantage of this variant is that winners always exist, while the disadvantage 
is that they may not to be unique. 
In this paper, we concentrate on voting systems with \emph{strict dominance} and 
the \emph{unique-winner condition}. 
It is not hard to translate all our results to the variants with weak dominance and 
possibly multiple winners.

\subsection{Range Voting}\label{SubSec22}

Approval voting can be regarded as a specific instance of a \emph{range voting} 
system. 
(Unfortunately, the terminology is not unique in the literature on voting and
instead of range voting various other names are used, e.g., scoring-based
voting, average voting, utility voting, and ratings summation.) \ 
Elections of such voting systems
are specified as triples $(C,V,(s_v)_{v \myin V})$, 
where each \emph{scoring function} $s_v : C \to \NAT$ specifies how many 
points voter $v$ gives to each of candidates, for all $v \myin V$. 
A candidate with strictly more points than all other candidates is defined as the winner. 
Therefore, candidate $c^*$ wins the range election $(C,V,(s_v)_{v \myin V})$ iff 
for all $c \myin C \setminus \{c^*\}$ it holds
\begin{equation}
\sum_{v \myin V} s_v(c^*) > \sum_{v \myin V} s_v(c).
\label{WINSCO}
\end{equation}
%

Another well-known voting system is the \emph{Borda voting} system, also 
called \emph{Borda count} and developed already in the 18th century by 
the French mathematician and political scientist J.-C. de Borda.
Each single voter ranks all candidates from top to bottom without ties, i.e., in a 
strictly decreasing manner, under the simplest form of the Borda count by giving 
$|C|-1$ points to the top candidate, $|C|-2$ points to the next one and so on,
with $0$ points for the candidate being ranked last.
The candidate with the most points is the winner.
With regard to the specification of the voters' preferences and the winner 
only, the Borda count can be interpreted as a specific instance of range 
voting, where all scoring functions $s_v$ are injective and fulfill the 
range restriction $s_v(C) = \{0,1,\ldots,|C|-1\}$. 
But this similarity is rather simplistic.
If additional features and properties are considered, then range voting and
the Borda count show strong differences and these prevent such an 
interpretation in many cases.
In view of the present paper it is important that such an interpretation 
makes it impossible to control Borda elections by changing the set of 
candidates -- procedures, which are studied in the literature, e.g., in
\cite{Russell}, \cite{Eklind} and \cite{Lorregia} -- since a change of 
$C$ usually destroys the the range restriction
$s_v(C) = \{0,1,\ldots,|C|-1\}$.

\subsection{Preference-Based Voting}\label{SubSec23}

The four voting systems we consider in the remainder of this section are 
\emph{preference-based}. 
This means that, as in the case of Borda voting, each single voter ranks all 
candidates from top 
to bottom in a strictly decreasing manner.
In contrast with Borda voting, however, now the individual preferences of 
the single voters $v$ 
are modeled by means of linear strict orders $>_v$ (that is, asymmetric 
and transitive relations,
where each pair of different elements is comparable) on the set $C$, which
directly describes the strictly decreasing order of the candidates. 
For a given election $(C,V,(>_v)_{v \myin V})$ with a so-called 
\emph{preference profile} 
$(>_v)_{v \myin V}$ the four following preference-based voting systems only 
differ in their 
voting rules.

In the \emph{Condorcet voting} system (named after the 18th-century French mathematician and 
philosopher N.\ de Condorcet) candidate $c$ strictly dominates another candidate $d$ 
iff the number of voters $v$ with $c >_v d$ is strictly larger than the number of 
voters $v$ with $d >_v c$. 
As a consequence, candidate $c^*$ wins the Condorcet election $(C,V,(>_v)_{v \myin V})$ iff 
\begin{equation}
|\{v \myin V : c^* >_v c\}| > |\{v \myin V : c >_v c^*\}|,
\label{WINCON}
\end{equation}
for all $c \myin C \setminus \{c^*\}$.
Already Condorcet noted a voting paradox that nowadays is called \emph{the Condorcet 
paradox}.
In our terminology it means that the strict dominance relation of a Condorcet
election may contain cycles -- even if it relates each pair of different candidates
(i.e., is a so-called \emph{tournament relation}).
In such a case it may happen that there exists no winner.

When the \emph{plurality voting} system is used, the most common voting system in
the Anglo-Saxon world, then candidate $c$ strictly dominates another candidate 
$d$ iff the number of voters with $c$ as top preference 
is strictly larger than the number of voters with $d$ as top preference. 
Thus, candidate $c^*$ wins the plurality election $(C,V,(>_v)_{v \myin V})$ iff 
\begin{equation}
\mbox{$
|\{v \myin V : c^* = \max_v C\}| > |\{v \myin V : c = \max_v C\}|,
$}
\label{WINPLU}
\end{equation}
for all $c \myin C \setminus \{c^*\}$.
In (\ref{WINPLU}) $\max_v C$ denotes the greatest element of the set $C$ w.r.t. the 
linear strict order $>_v$, i.e., that element $c \myin C$ for which $c >_v d$ for 
all $d \myin C \setminus \{c\}$.

The \emph{maximin voting} system uses the maximin principle, originally formulated 
for two player zero-sum games, and also defines the winner by means of the 
cardi\-na\-li\-ties of the sets $\{v \myin V : c >_v d\}$. 
If we call $|\{v \myin V : c >_v d\}|$ the \emph{advantage} of candidate $c$ 
over candidate $d$ and define the function
\[
\Phi : C \to \NAT
\qquad\qquad
\Phi(c) = \min \{ |\{v \myin V : c >_v d\}| : d \myin C \setminus \{c\} \}
\]
that yields for each candidate the minimum of all its advantages over all other 
candidates, then candidate $c^*$ wins the maximin election $(C,V,(>_v)_{v \myin V})$ iff 
\begin{equation}
\Phi(c^*) > \Phi(c),
\label{WINMAX}
\end{equation}
for all $c \myin C \setminus \{c^*\}$,
that is, iff it maximizes the minimum of all advantages over all other candidates
and this maximum advantage is unique.

Finally, we consider the \emph{Bucklin voting} system, named after the 
American J.W.\ Bucklin but already proposed by Condorcet. 
Strictly speaking, we describe a rule that is known as simplified Bucklin
rule.
This rule is based on the candidates' \emph{Bucklin scores}, which are computed via 
the function
\[
\Psi : C \to \NAT_{>0}
\qquad\qquad
\Psi(c) = \min \{k \myin \NAT_{>0} : 
                 |\{v \myin V : c \myin \Rank_{v,k}\}| > \frac{n}{2}\},
\]
where $\Rank_{v,k} := \{c \myin C : |\{d \myin C : d >_v c\}| < k \}$ is 
the set of candidates which are ranked among the top $k$ positions by voter 
$v$, for all $k \myin \NAT_{>0}$ and $v \myin V$. 
In words, the Bucklin score $\Psi(c)$ of candidate $c$ is the least (positive) 
natural number $k$ such that $c$ is ranked among the top $k$ positions by 
(strictly) more than half of the voters. 
By definition then, candidate $c^*$ wins the Bucklin election 
$(C,V,(>_v)_{v \myin V})$ iff 
\begin{equation}
\Phi(c^*) < \Phi(c),
\label{WINBUC}
\end{equation}
for all $c \myin C \setminus \{c^*\}$, that is, iff it minimizes the Bucklin 
scores and this minimum is unique.

\section{Control Problems in Voting Systems}\label{Sec3} 

This section introduces the different types of control we consider in this paper
and presents their computational complexities for the voting systems we have 
introduced in the previous section.

\subsection{Constructive and Destructive Control by Deleting}\label{SubSec31} %

If control problems in voting systems are modeled mathematically, then it is 
assumed that the authority conducting the election (the actor mentioned in
the introduction, in the literature on voting systems is usually called the 
\emph{chair}) knows all individual preferences of the single voters. 
His goal then is to achieve a specific result by a strategic change of 
the set of candidates or voters, respectively, but not of the individual 
preferences of the voters. 
To conceal his manipulations, the chair furthermore tries to change these sets as 
little as possible.

The literature on voting systems investigates several types of strategic changes.
In the present paper, we allow only \emph{deleting candidates and voters}, 
respectively, as the chair's possibilities. 
We mainly focus on \emph{constructive control} as investigated 
for the first time by \cite{BarTovTri} in view of computational complexity. 
Using this type of control, the chair's goal is to make his favourite candidate 
$c^*$ the winner. 
The counterpart of constructive control is \emph{destructive control}.
Here the chair tries to prevent a specific disliked candidate $c^*$ from being 
the winner. 
First results on the computational complexity of this type of control were given by 
\cite{HemHemRot}. 

The control of elections by deleting candidates or voters
can be stated as a minimization problem (\cite{BarTovTri,HemHemRot}). 
If constructive control is done by deleting candidates, then the problem is as 
follows: 
Given an election $(C,V,I)$ and the specific candidate $c^*$, compute a 
minimum set of candidates $M$ such that $c^* \in C \setminus M$ and the deletion 
of $M$ from $C$ and of its candidates from the individual preferences ensures that
$c^*$ is the winner of the resulting election.
To allow for an easier modeling in Section~\ref{Sec4}, we consider the dual 
maximization-problem and ask for
\begin{itemize}
\item[(a)] a maximum subset $C^* \subseteq C$ such that $c^* \in C^*$ and
           $c^*$ wins the election $(C^*,V,I)$, in which the 
           original individual preferences are restricted to $C^*$.
\end{itemize}
It is obvious that from the set $C^*$ then the desired set $M$ is obtained by 
defining $M := C \setminus C^*$. 
In an analogous manner we specify the constructive control problem by deleting
voters as maximization-problem for a given election $(C,V,I)$ 
and the specific candidate $c^*$. 
Again we ask for
\begin{itemize}
\item[(b)] a maximum subset $V^*$ of $V$ such that $c^*$ wins the election 
           $(C,V^*,I)$, in which the original individual preferences are 
           restricted to $V^*$.
\end{itemize}
Using the specifications (a) and (b), we can immediately obtain specifications of the 
destructive variants of the controls via deleting by replacing the phrase `such 
that $c^*$ wins' by the phrase `such that $c^*$ does not win'.

In Section~\ref{Sec2} we have explained by means of approval voting and the Condorcet
paradox that in voting systems with strict dominance and the unique-winner condition 
it may happen that no candidate wins.
This implies that also solutions of the control problems do not necessarily have 
to exist.
When we later model control problems as integer programs, then the non-existence 
of a winner of a control problem will be expressed by the fact that the modeling 
program has no feasible solution. 

\subsection{Complexity of Control by Deleting Candidates or Voters}\label{SubSec32} 

Given a voting system, some control problems may be easy, some may be hard, and 
in some cases it may even be impossible for the chair to reach his goal. 
If a control problem is easy, one says that the voting system is \emph{vulnerable} 
to this type of control. In this case, there exists an efficient algorithm that solves the problem 
to optimality in polynomial time. If it is hard, one says that it is \emph{resistant} to 
this type of control. This is formally specified by the NP-hardness of the decision 
problem corresponding to the original optimization problem with a bound for the 
size as an additional input. In case of constructive control by deleting candidates, an instance
of the decision problem corresponding to the original minimization problem consists 
of an election $(C,V,I)$, the specific candidate $c^*$, and a natural number 
$k$. The question is whether it is possible to delete at most $k$ candidates 
such that $c^*$ wins the resulting election. If a problem is unsolvable, one 
says that it is \emph{immune} to this type of control. This means that it is never possible 
for the chair to reach his goal by the corresponding control action. 
In other words, no feasible solution exists for the unsolvable control problem.

Inspired by the seminal paper of \cite{BarTovTri}, scientists have investigated 
the hardness of control problems via the methods of complexity theory. 
See, e.g., the references given in the introduction or by \cite{BraConEnd} 
in Section 3.2 of their study.
In the following, we summarize the results concerning the voting systems we 
have discussed in Section~\ref{Sec2} and the four types of control we have 
considered above.

Approval voting and Condorcet voting are vulnerable to destructive control by 
deleting voters and to constructive control by deleting candidates, resistant 
to constructive control by deleting voters, and immune to destructive control 
by deleting candidates. 
For the constructive control types these results are proved by 
\cite{HemHemRot} for approval voting and 
by \cite{BarTovTri} for Condorcet voting;
for the destructive control types they are 
proved by \cite{HemHemRot}. 
Since we have introduced approval voting as a 
specific instance of range voting, also the latter kind of voting is 
resistant to constructive control by deleting voters and immune to 
destructive control by deleting candidates. 
The Borda voting system is proved to be vulnerable to destructive 
control by deleting voters. 
The questions on complexity of constructive control types, as well as 
destructive control by deleting candidates are still open to the 
best of our knowledge. 
However, for elections of precisely three candidates the Borda 
voting system is vulnerable to constructive control by deleting voters, 
as shown by \cite{Russell}. 
We refer to the study of 
\cite{Russell}, \cite{Eklind} and \cite{Lorregia} 
for further discussions 
concerning the complexity of control of Borda elections. 
Plurality voting is vulnerable to 
constructive as well as destructive control by deleting voters and resistant 
to constructive as well as destructive control by deleting candidates. Here 
the proofs for the constructive control types are presented by \cite{
BarTovTri} and those for the destructive control types again by \cite{
HemHemRot}. For maximin voting the 
situation is exactly contrary to plurality voting. The maximin voting system 
is vulnerable to constructive as well as destructive control by deleting 
candidates and resistant to constructive as well as destructive control by 
deleting voters. Concerning proofs of these facts we refer to the study of \cite{FalHemHem}. 
Finally, Bucklin voting is vulnerable to destructive control by deleting 
voters and resistant to the three other types of control, i.e., destructive 
control by deleting candidates and constructive control by deleting 
candidates as well as voters. These facts are shown by \cite{ErdFelRotSch}.

\section{Modeling Control Problems as Integer Programs}\label{Sec4} 
Using \emph{Linear Programming} (LP) has been shown to be successful for optimization problems in various fields (\cite{Chv}). 
In the so-called \emph{standard form}, an LP formulation consists of a linear 
\emph{objective function} $f : \REAL^n_{\geq 0} \to \REAL_{\geq 0}$ given as
$$f(x_1,\ldots,x_n) = \sum_{j=1}^n c_j x_j$$ that has to be maximized, and $m$ linear 
\emph{inequality constraints}
$$\sum_{j=1}^n a_{i,j} x_j \leq b_i, 1 \leq i \leq m.$$ Furthermore, there is the \emph{non-negative variables condition} requiring that $x_j \in \mathds{R}_{\geq 0}$.

In many practical applications it is additionally required that the variables 
range over the set $\NAT$ only. 
Restricting the variables to only non-negative integer values leads to an \emph{integer linear programming} (ILP) formulation. 
In contrast to LP, the ILP problem is NP-hard as shown by \cite{Karp} even for 
the special case of \emph{binary integer programming} (abbreviated as BIP), where $x_j\ \in \{0,1\}$ is required. 
Nevertheless, there are tools available that also allow to solve larger 
instances of ILP and BIP by techniques like relaxation and branch-and-bound. 
Examples are the \textsc{Mathlab} LP solver, Xpress, Gurobi and the \textsc{Cplex} tool of 
IBM.

In this section, we demonstrate how the hard control 
problems of the voting systems introduced in Section~\ref{Sec3} can be specified as 
ILPs and BIPs, 
respectively. Without loss of generality, we work under the 
assumption that  $c^* = c_1$, i.e., that the chair's goal is 
having the first candidate winning (loosing) for  constructive (destructive) control. 
We consider the control problems as maximization problems 
as introduced in Section~\ref{Sec3} via the specifications (a) and (b).
We restrict us to constructive control and sketch
in Section~\ref{SubSec46} how our models for constructive control can be 
adapted for destructive one.

\subsection{Range Election Model}\label{SubSec41}

Since approval voting is the specific case of range voting where all
scores are zero or one, we start 
our modeling with the constructive control of range voting by deleting 
voters. 
To this end, we assume a range election $(C,V,(s_v)_{v \myin V})$ to 
be given. 
As a first step, we combine the list of scoring functions $(s_v)_{v \in V}$ 
into a single matrix $A \myin \NAT^{m \times n}$ such that 
\[
A_{ij} = s_{v_j}(c_i),
\]
for all $i \myin \{1,\ldots,m\}$ and 
$j \myin \{1,\ldots,n\}$. 
Next, we represent a solution of the control problem by the binary decision 
vector $x \in \{0,1\}^n$ such that $x_j = 1$ iff voter $v_j$ is allowed 
to vote, for all $j \in \{1,\ldots,n\}$. 
Consequently, we arrive at the binary integer program (\SBE) of
Figure \ref{BIPRANGE} that models the given problem.
\begin{figure}[t]
\begin{flalign}
~~~\mbox{max} \; & \displaystyle\sum_{j=1}^n x_j & \label{eq:sbe0}
\\
\mbox{s.t.}\; & \displaystyle\sum_{j=1}^n A_{1j}x_j - \displaystyle\sum_{j=1}^n A_{ij}x_j \geq 1
                                                & i \in \{2, \ldots, m\} \label{eq:sbe1}
\\
              & x_j \in \left\{0,1\right\}    & j \in \{1, \ldots, n\} \label{eq:sbe2}
\end{flalign}
\caption{Binary Integer Program for Range Elections (Deleting Voters)}
\label{BIPRANGE}
\end{figure}

Because we ask for a maximum subset $V^*$ of $V$ such that $c_1$ wins the 
range election $(C,V^*,(s_v)_{v \in V^*})$, (\ref{eq:sbe0}) 
describes the objective function as the maximum number of voters allowed to 
take part in voting. 
The set of constraints (\ref{eq:sbe1}) supposes 
that candidate $c_1$ is the unique winner of $(C,V^*,(s_v)_{v \in V^*})$ since
it collects the largest total amount of scores. 
Finally, (\ref{eq:sbe2}) states that variables $x_1, \ldots, x_n$ are binary. 
Note that the solution of the proposed ({\SBE}) program is in the dual form 
respecting the initial problem statement given in Section~\ref{SubSec22}. 
In fact, each value $x_j$, where $j \in \{1,\ldots,n\}$, with $x_j = 0$ of the solution 
vector $x$ defines voter $v_j$ to be excluded from the voting process in 
the standard form.

We have already mentioned that a solution of a control problem not necessarily 
has to exist and this is expressed by the fact that the modeling integer 
program has no feasible solution. 
The sufficient condition of the existence of a feasible solution for the
program (\SBE) is the existence of at least one voter whose preference 
list quotes $c_1$ as the best candidate. 
Otherwise, deleting any subset of voters may not lead to a feasible solution 
where $c_1$ wins.

\subsection{Condorcet Election Model}\label{SubSec42}

As the second problem, we investigate the constructive control by deleting voters 
for a given Condorcet election $(C,V,(>_v)_{v \in V})$. 
Doing so, we represent the preference profile $(>_v)_{v \myin V}$ by a single 
binary matrix $A \myin \{0,1\}^{(m-1) \times n}$ such that 
for all $i \myin \{1,\ldots,m-1\}$ and $j \myin \{1,\ldots,n\}$ it holds
\[
A_{ij} = 1 \iff c_1 >_{v_j} c_{i+1}.
\]
To give an example, if we assume the set $C = \{c_1,c_2,c_3,c_4\}$ of candidates,
the set $V = \{v_1,v_2,v_3\}$ of voters, and the preference profile 
\[
\begin{array}{l@{\qquad}l}
v_1: & c_1 >_{v_1} c_2 >_{v_1} c_3 >_{v_1} c_4 \\
v_2: & c_1 >_{v_2} c_3 >_{v_2} c_2 >_{v_2} c_4 \\
v_3: & c_4 >_{v_3} c_3 >_{v_3} c_2 >_{v_3} c_1,
\end{array}
\]
then the binary matrix $A \myin \{0,1\}^{3 \times 3}$ that represents
this preference profile looks as follows:
\[
A = \left ( \begin{array}{cccc}
            1 & 1 & 0 \\
            1 & 1 & 0 \\
            1 & 1 & 0  
            \end{array}
    \right )
\]

We establish the binary decision vector $x\in \{0,1\}^n$ to represent the 
solution $V^*$ of the problem, that is, have $x_j=1$ iff voter $v_j$ is 
permitted to vote in $(C,V^*,(>_v)_{v \in V^*})$, for all $j \in \{1,\ldots,n\}$,
and candidate $c_1$ is the 
unique winner of this election.
Thus, we arrive at the binary integer program (\CE) of
Figure \ref{BIPCONDORCET} for Condorcet elections.
\begin{figure}[t]
\begin{flalign}
~~~\mbox{max}\;  & \displaystyle\sum_{j=1}^n x_j & \label{eq:ce0}
\\
\mbox{s.t.}\; & \displaystyle\sum_{j=1}^n \left( 2A_{ij}-1\right) x_j \geq 1
                                                & i \in \{1, \ldots, m-1\} \label{eq:ce1}
\\
              & x_j \in \left\{0,1\right\}    & j \in \{1, \ldots, n\} \label{eq:ce2}
\end{flalign}
\caption{Binary Integer Program for Condorcet Elections (Deleting Voters)}
\label{BIPCONDORCET}
\end{figure}

Constraint (\ref{eq:ce0}) defines the objective function as the maximum number of 
voters allowed to take part in voting. 
Constraints (\ref{eq:ce1}) ensure that for all $i \in \{2,\ldots,m\}$ the 
number of voters who gives a vote to candidate $c_1$ over candidate 
$c_i$ is strictly greater than the number of those who prefers $c_i$ over $c_1$. 
Indeed, the form of (\ref{eq:ce1}) is equivalent to the form
\[
\displaystyle\sum_{j=1}^n A_{ij} x_j > \displaystyle\sum_{j=1}^n \left(1 - A_{ij}\right) x_j, 
\]
which is based on the idea that for each voter $v_j$ either $c_1$ dominates $c_i$ and $A_{ij}=1$, 
or $c_1$ is dominated by $c_i$, and therefore $A_{ij}=0$ and the 
coefficient $\left(1-A_{ij}\right)$ is 1. 
In fact, this set of constraints makes $c_1$ a unique winner. 
Finally, (\ref{eq:ce2}) states that variables $x_1, \ldots, x_n$ are binary. 

For the program (\CE) the necessary condition of the existence of a feasible 
solution requires that for each candidate $c_i$, $i \in \{2,\ldots,m\}$, 
there exists at least one voter $v_j$, $j \in \{1,\ldots,n\}$, who prefers 
$c_1$ over $c_i$. 
A feasible solution always exists when the sufficient condition holds, 
thus, when there exists at least one voter who gives a top preference 
to candidate $c_1$ over any other candidates.

\subsection{Plurality Election Model}\label{SubSec43}

As the third problem, we consider the constructive control by deleting 
candidates for a given plurality election $(C,V,(>_v)_{v \myin V})$. 
Here, we assume the election's preference profile $(>_v)_{v \myin V}$ to be 
specified by a list of $n$ binary matrices 
$A^1,\ldots,A^n \in \{0,1\}^{m \times m}$ such that 
for all $i,k \in \{1,\ldots,m\}$ and $j \in \{1,\ldots,n\}$ it holds
\[
A^j_{ik} = 1 \iff c_i >_{v_j} c_k.
\]
Note, that each $A^j$ is nothing else than the binary matrix representation of 
the linear strict order $>_{v_j}$.
Hence, in the case of the example from Section~\ref{SubSec42} we get the following
binary matrices $A^1, A^2, A^3 \myin \{0,1\}^{4 \times 4}$:
\[
A^1 = \left ( \begin{array}{cccc}
            0 & 1 & 1 & 1 \\
            0 & 0 & 1 & 1 \\
            0 & 0 & 0 & 1 \\
            0 & 0 & 0 & 0 
            \end{array}
      \right )
\quad
A^2 = \left ( \begin{array}{cccc}
            0 & 1 & 1 & 1 \\
            0 & 0 & 0 & 1 \\
            0 & 1 & 0 & 1 \\
            0 & 0 & 0 & 0 
            \end{array}
      \right )
\quad
A^3 = \left ( \begin{array}{cccc}
            0 & 0 & 0 & 0 \\
            1 & 0 & 0 & 0 \\
            1 & 1 & 0 & 0 \\
            1 & 1 & 1 & 0 
            \end{array}
      \right )
\quad
\]

Since we seek for a maximum subset $C^*$ of the set $C$ of candidates such 
that $c_1$ wins the plurality election $(C^*,V,(>^*_v)_{v \myin V})$, where 
$(>^*_v)_{v \myin V}$ denotes the restriction of the preference profile 
$(>_v)_{v \myin V}$ to the set $C^*$, we represent the solution by a binary 
decision vector $x \in \{0,1\}^m$ such that $x_i = 1$ iff candidate $c_i$
is admitted to take part in the election, for all $i \in \{1,\ldots,m\}$.
Furthermore, we introduce a set of auxiliary binary variables $z^j_i$, where 
$z^j_i=1$ holds iff candidate $c_i$ is of the highest preference for voter 
$v_j$ among the set of candidates chosen by the vector $x$, for all
$i \in \{1,\ldots,m\}$ and $j \in \{1,\ldots,n\}$.
In such a way, the solution of the problem can be derived by the 
binary integer program (\PE) of Figure \ref{BIPPLURALITY}.
\begin{figure}[t]
\begin{flalign}
~~~\mbox{max} \; & \displaystyle\sum_{i=1}^m x_i 
                & \label{eq:pv0}
\\
\mbox{s.t.}\; & \displaystyle\sum_{k=1}^m A^j_{ki} x_k+m z^j_i\leq m 
                & i \in \{1, \ldots, m\}, \; j \in \{1, \ldots n\} \label{eq:pv1}
\\
& \displaystyle\sum_{k=1}^m A^j_{ki} x_k+ z^j_i - x_i \geq 0 
                & i \in \{1, \ldots, m\}, \; j \in \{1, \ldots n\} \label{eq:pv2}
\\
& \displaystyle\sum_{j=1}^n z^j_i \leq n x_i 
                & i \in \{2, \ldots, m\} \label{eq:pv3}
\\
& \displaystyle\sum_{j=1}^n z^j_1 - \displaystyle\sum_{j=1}^n z^j_i \geq 1 
                & i \in \{2, \ldots, m\} \label{eq:pv4}
\\
& x_i \in \left\{0,1\right\} & i \in \{1, \ldots, m\} \label{eq:pv5}
\\
& z^j_i \in \left\{0,1\right\} & i \in \{1, \ldots, m\}, \; j \in \{1, \ldots, n\} \label{eq:pv6}
\end{flalign}
\caption{Binary Integer Program for Plurality Elections (Deleting Candidates)}
\label{BIPPLURALITY}
\end{figure}

Here, (\ref{eq:pv0}) defines the objective function as the maximum number 
of candidates admitted to take part in the election. 
Constraints (\ref{eq:pv1}) imply that for all $i \in \{1,\ldots,m\}$ and 
$j \in \{1,\ldots,n\}$ voter $v_j$ may give the highest preference to candidate 
$c_i$ over all other candidates selected by the vector $x$ only if 
there exists no candidate $c_k$, $k \in \{1,\ldots,m\}$,
that is preferred by $v_j$ over $c_i$. 
In its turn, (\ref{eq:pv2}) imposes that when candidate $c_i$ is 
allowed to the contest, either $c_i$ must be of the highest preference for the 
voter $v_j$, or a candidate $c_k$ preferred over $c_i$ must exist. 
Each of constraints (\ref{eq:pv3}) enforces that candidate $c_i$ cannot be 
of the highest preference for any voter if he is not permitted to 
participate in the election, for all $i \in \{2,\ldots,m\}$.
Thus, the constraint requires 
for all $i \in \{2,\ldots,m\}$ and $j \in \{1,\ldots,n\}$ that
$z^j_i=0$ if candidate $c_i$ has been deleted from the election. 
The next constraints (\ref{eq:pv4}) ensure that candidate $c_1$ is the 
candidate with the highest number of voters having $c_1$ as top priority, 
and therefore is strictly preferred over other candidates. 
Implicitly, this set of constraints requires $x_1=1$ and hence candidate 
$c_1$ has to be selected in any feasible solution. 
Finally, (\ref{eq:pv5}) and (\ref{eq:pv6}) state that variables 
$x_1,\ldots,x_m$ and $z^1_1,\ldots,z^n_m$ are binary. 

There always exists a feasible solution 
for the program (\PE), which is guaranteed by possible deleting of all 
candidates but candidate $c_1$ as the worst case.

\subsection{Maximin Election Model}\label{SubSec44}

The fourth problem we consider is the constructive control by deleting voters 
in elections with the maximin voting rule. 
We assume that the preference profile $(>_v)_{v \in V}$ is given 
by a list of $n$ binary matrices $A^1,\ldots,A^n \in \{0,1\}^{m \times m}$ 
such that 
\[
A^j_{ik} = 1 \iff c_i >_{v_j} c_k,
\]
for all $i,k \in \{1,\ldots,m\}$ and $j \in \{1,\ldots,n\}$. 
We use a binary decision vector $x \in \{0,1\}^n$ to represent the 
solution, where $x_j = 1$ iff voter $v_j$ has a permission to vote, for all 
$j \in \{1,\ldots,n\}$.
Thus, the advantage of candidate $c_i$ over candidate $c_k$ can be computed as 
\[
adv_{maximin}\left(c_i,c_k\right) = \displaystyle\sum_{j=1}^n A^j_{ik}x_j ,
\]
for all $i,k \in \{1,\ldots,m\}$. 
Now, let candidate $c_1$ be the winner of the maximin
election and let the positive 
integer variable $b$ define the minimum advantage of $c_1$ over any other 
candidate. 
Subsequently, let for all $i,k \in \{1,\ldots,m\}$ with $i\neq 1$ and $i\neq k$ 
the auxiliary binary variable $z_{ik}=1$ denote a situation when 
\[
adv_{maximin}\left(c_i,c_k\right) < b.
\]
Then, the solution of the posed problem can be derived by the integer 
program (\ME) of Figure \ref{IPMAXIMIN}.
\begin{figure}[t]
\begin{flalign}
~~~\mbox{max}\;  & \displaystyle\sum_{j=1}^n x_j &\label{eq:mv0}
\\
\mbox{s.t.}\; & \displaystyle\sum_{j=1}^n A^j_{ik}x_j - n\left(1-z_{ik}\right) \leq b-1 
              & i \in \{2, \ldots m\}, \; k \in \{1, \ldots m\}, \; i\neq k \label{eq:mv1}
\\
& \displaystyle\sum_{k=1, \; k\neq i}^m z_{ik} \geq 1 
              & i \in \{2, \ldots m\} \label{eq:mv2}
\\
& b \leq \displaystyle\sum_{j=1}^n A^j_{1k} x_j 
              & k \in \{2, \ldots, m\} \label{eq:mv3}
\\
& x_j \in \left\{0,1\right\} 
              & j \in \{1, \ldots, n\} \label{eq:mv4}
\\
& z_{ik} \in \left\{0,1\right\} 
              & i \in \{2, \ldots m\}, \; k \in \{1, \ldots m\}, \; i\neq k \label{eq:mv5}
\\
& b \in \mathbb{N}_{>0} \label{eq:mv6}
\end{flalign}
\vspace{-3mm}
\caption{Integer Program for Maximin Elections (Deleting Voters)}
\label{IPMAXIMIN}
\end{figure}

Here, (\ref{eq:mv0}) defines the objective function as the maximum 
number of voters allowed to take part in voting. 
Next, the set of constraints (\ref{eq:mv1}) strictly bound the 
advantage values of any candidate but $c_1$ by $b$. 
Its combination with constraints (\ref{eq:mv2}) forces at least the minimal advantage 
value of each candidate $c_i$ to be bounded by $b$, for all $i \in \{2, \ldots m\}$.
In its turn, constraints (\ref{eq:mv3}) bound, and therefore define $b$ 
as the minimal advantage of $c_1$. 
Finally, (\ref{eq:mv4}) and (\ref{eq:mv5}) state that variables
$x_1,\ldots,x_n$ and $z_{21},\ldots,z_{mm}$ are binary, while (\ref{eq:mv6}) states that 
$b$ is a positive integer.

There always exists a feasible solution for the program (\ME) if candidate 
$c_1$ is a winner concerning at least one of the voters. 
In fact, if there is voter $v_j$, $j \in \{1,\ldots,n\}$, whose top preference 
is $c_1$, then deleting of all other voters in the worst case makes $c_1$ the winner 
of the maximin-based rule election.

\subsection{Bucklin Election Model}\label{SubSec45}

The last two constructive control problems address elections with the 
(simplified) Bucklin voting rule, which is resistant to the constructive 
control by deleting voters as well as by deleting candidates. 

We start first with the variant of the problem which stipulates deleting voters. 
Doing so, we assume now that the preference profile $(>_v)_{v \in V}$ is 
described by a list of $n$ binary matrices 
$A^1,\ldots,A^n \in \{0,1\}^{m \times m}$ such that for all
$i, k \in \{1,\ldots,m\}$ and $j \in \{1,\ldots,n\}$ it holds 
\[
A^j_{ik} = 1
\iff
\left \{ \begin{array}{l}
         k \in \{m',\ldots,m\} \mbox{ and voter $v_j$ ranks the} \\
         \mbox{candidate $c_i$ as $m'$-th in his preference list. }
         \end{array}
\right .
\]
In other words, candidate $c_i$ may obtain at least $m'$ as the 
personal score from voter $v_j$. 
In such a way, $c_i$ gets 1 for each entry in the row $i$ of binary matrix $A^j$ 
when it is the most preferred by $v_j$ over other candidates. 

In case of the example from Section~\ref{SubSec42} we get the following
binary matrices $A^1,\ldots,A^3 \in \{0,1\}^{4 \times 4}$:

\[
A^1 = \left ( \begin{array}{cccc}
            1 & 1 & 1 & 1 \\
            0 & 1 & 1 & 1 \\
            0 & 0 & 1 & 1 \\
            0 & 0 & 0 & 1 
            \end{array}
      \right )
\quad
A^2 = \left ( \begin{array}{cccc}
            1 & 1 & 1 & 1 \\
            0 & 0 & 1 & 1 \\
            0 & 1 & 1 & 1 \\
            0 & 0 & 0 & 1 
            \end{array}
      \right )
\quad
A^3 = \left ( \begin{array}{cccc}
            0 & 0 & 0 & 1 \\
            0 & 0 & 1 & 1 \\
            0 & 1 & 1 & 1 \\
            1 & 1 & 1 & 1 
            \end{array}
      \right )
\quad
\]

Subsequently, to represent the problem's solution we employ a binary 
decision vector $x \in \{0,1\}^n$, where $x_j = 1$ iff voter $v_j$ 
participates in voting. 
Therefore, $\sum_{j=1}^n A^j_{ik}x_j$ determines the number of voters 
ready to give the score $k$ to candidate $c_i$. 
Let, for all $i, k \in \{1,\ldots,m\}$, the auxiliary binary variable 
$z_{ik}=1$ describe the situation when strictly more than a half of 
the voters allowed to vote agree to give the score of $k$, i.e., rank $c_i$ 
among the top $k$ candidates.
Then, the solution of the problem can be obtained by the 
binary integer program (\BEV) of Figure \ref{BIPBUCKLIN}.
\begin{figure}[t]
\begin{flalign}
~~~\mbox{max}\;& \displaystyle\sum_{j=1}^n x_j &\label{eq:bev0}
\\
\mbox{s.t.}\; & \displaystyle\sum_{j=1}^n \left(\frac{1}{2}-A^j_{ik}\right) x_j + nz_{ik} \leq n-\frac{1}{2} 
            & i,k \in \{1, \ldots m\} \label{eq:bev1}
\\
& \displaystyle\sum_{j=1}^n \left(\frac{1}{2}-A^j_{ik}\right) x_j + nz_{ik} \geq 0 
            & i\in \{2, \ldots m\}, \; k \in \{1, \ldots m\} \label{eq:bev2}
\\
& \displaystyle\sum_{l=1}^m lz_{1l} + \left(n-k\right)z_{ik} \leq n-1 
            & i\in \{2, \ldots m\}, \; k \in \{1, \ldots m\} \label{eq:bev3}
\\
& \displaystyle\sum_{l=1}^m z_{1l} \geq 1 
            & \label{eq:bev4}
\\
& x_j \in \left\{0,1\right\} 
            & j \in \{1, \ldots, n\} \label{eq:bev5}
\\
& z_{ik} \in \left\{0,1\right\} 
            & i \in \{1, \ldots m\}, \; k \in \{1, \ldots m\} \label{eq:bev6}
\end{flalign}
\caption{Binary Integer Program for Bucklin Elections (Deleting Voters)}
\label{BIPBUCKLIN}
\end{figure}
 
Constraint (\ref{eq:bev0}) defines the objective function as the maximum number of 
voters allowed to take part in voting. 
Constraints (\ref{eq:bev1}) imply that each candidate $c_i$ may earn the 
score $k$ iff it obtains votes of strictly more than a half of the
participating voters, for all $i,k \in \{1, \ldots m\}$.
Specifically, the form of (\ref{eq:bev1}) is the reduction of the form
\[
\frac{1}{2} + \frac{1}{2} \displaystyle\sum_{j=1}^n x_j - \displaystyle\sum_{j=1}^n A^j_{ik} x_j + nz_{ik} 
\leq 
n,
\]
where the first constant ensures the strictness concerning the half of the total 
number of votes given to $c_i$, 
the second term defines the half of all the available votes, the third term 
calculates the number of participating voters 
ready to give the score $k$ to $c_i$, and finally the combination of the 
fourth term and the constant in the right hand side introduces 
trigger variable $z_{ik}$ to handle the corresponding situation. 
Next, for all $i \in \{2,\ldots,m\}$ and $k \in \{1,\ldots,m\}$, constraints
(\ref{eq:bev2}) force indicator variable $z_{ik}$ be equal to 1 every 
time when $c_i$ earns votes of more than a half of the participating voters. 
The form of (\ref{eq:bev2}) is in fact equivalent to the inequality
\[
\frac{1}{2} \displaystyle\sum_{j=1}^n x_j - \displaystyle\sum_{j=1}^n A^j_{ik} x_j + nz_{ik} 
\geq 
0.
\]
\noindent 
When each of the variables $z_{ik}$ reveals that the given threshold is reached, 
then (\ref{eq:bev3}) 
requires the minimal score obtained by the first candidate $c_1$ be strictly 
less than the scores obtained by any other candidates. 
Indeed, the form of constraints (\ref{eq:bev3}) results from the form
\begin{equation}
\displaystyle\sum_{l=1}^m lz_{1l} +1 \leq kz_{ik} + n \left(1-z_{ik}\right),\label{eq:bevNote}
\end{equation}
where the left hand side defines the minimal score obtained by $c_1$ and 
ensures the strictness of the inequality 
of the voting rule, while the right hand side defines the score obtained 
by $c_i$ and bounds the former. Note that constraints (\ref{eq:bev2}) are imposed for all the candidates but the first one. 
Therefore, variable $z_{1k}$ may not take the value of 1 for multiple scores $k$, $k=1 \ldots m$, while variable $z_{ik}$ for $i=2 \ldots m$ must do this according to (\ref{eq:bev2}).
This makes only one of the variables $z_{1l}$ equal to 1 in the sum in the left hand side part of (\ref{eq:bevNote}). It is exactly the case corresponding to one of the possible scores $l$ (at least the smallest one), since the left hand side part is bounded by the right hand side part, while constraint (\ref{eq:bev4}) asks at least one variable $z_{1k}$ related to 
$c_1$ be set to 1. 
In fact, (\ref{eq:bev4}) guarantees that at least the indicator pointing to 
the least score obtained by $c_1$ will trigger. 
Finally, (\ref{eq:bev5}) and (\ref{eq:bev6}) state that variables $x_1,\ldots,x_n$ 
and $z_{11},\ldots,z_{mm}$ are binary.

Similarly to the program (\CE), the same necessary condition on the existence 
of a feasible solution must hold for the program (\BEV).
It requires the existence of at least one voter who prefers $c_1$ over 
$c_i$, for each $i \in \{2,\ldots,m\}$. If this condition 
fails, then no subset of voters can be deleted in order to guarantee 
$c_1$'s win. As the sufficient condition for the feasible 
solution, it is required that at least one voter exists, who 
gives a top preference to $c_1$ over any other candidates.

Herein, we deal with the second variant of the control problem, where 
a subset of candidates may be deleted from the Bucklin election. 
Compared with the first case we change the input.
Now, we suppose that the preference profile $(>_v)_{v \in V}$ is 
modeled as in the cases of plurality voting and maximin voting, that is, by a list of $n$ binary 
matrices $A^1,\ldots,A^n \in \{0,1\}^{m \times m}$ such that 
\[
A^j_{ik} = 1 \iff c_i >_{v_j} c_k,
\]
for all $i,k \in \{1,\ldots,m\}$, and $j \in \{1,\ldots,n\}$. 
A binary decision vector $x \in \{0,1\}^m$ is used for the solution description, 
where $x_i = 1$ iff candidate $c_i$ is admitted to take part in the election, 
for all $i \in \{1,\ldots,m\}$. 

We construct the model in such a way that for all $i \in \{1,\ldots,m\}$ and 
$j \in \{1,\ldots,n\}$ candidate $c_i$ may obtain at 
least $m'$ as a personal score from voter $v_j$ when exactly $m'-1$ 
candidates have higher ranks in $v_j$'s preference list. 
To reveal this fact, we use a set of auxiliary binary variables 
$y^j_{il}$, where $j \in \{1, \ldots, n\}$ and $i,l \in \{1, \ldots, m\}$,
such that 
$y^j_{il}=1$ when candidate $c_i$ can get the score of value $l$ from voter $v_j$, i.e., when the number of available candidates preferred by 
$v_j$ over $c_i$ is strictly less than $l$. 
Subsequently, we determine the number of voters ready to give the score 
$l$ to candidate $c_i$ as $\sum_{j=1}^n y^j_{il}$. 
Let the further auxiliary binary variables $z_{il}=1$ for all $i, l \in \{1,\ldots,m\}$
denote the situation when strictly 
more than half of the voters rank candidate $c_i$ among the top most $l$ candidates. 
Then, the solution of the posed problem can be computed via the binary 
integer program (\BEC) of Figure \ref{BIPBUCKLINXX}.
\begin{figure}[t]
\begin{flalign}
~~~\mbox{max}\;& \displaystyle\sum_{i=1}^m x_i &\label{eq:bec0}
\\
\mbox{s.t.}\; & \displaystyle\sum_{k=1}^m A^j_{ki} x_k + \left(m+1\right)y^j_{il}\leq m+l & j \in \{1, \ldots, n\},\;i,l \in \{1, \ldots, m\} \label{eq:bec1}
\\
& \displaystyle\sum_{k=1}^m A^j_{ki} x_k + my^j_{il} - mx_i \geq l-m & j \in \{1, \ldots, n\},\;i,l \in \{1, \ldots, m\} \label{eq:bec2}
\\
& \displaystyle\sum_{j=1}^n \displaystyle\sum_{l=1}^m y^j_{il} - nmx_i \leq 0 & i\in \{2, \ldots, m\} \label{eq:bec3}
\\
& nz_{il} - \displaystyle\sum_{j=1}^n y^j_{il} \leq \frac{n-1}{2} & i,l\in \{1, \ldots, m\} \label{eq:bec4}
\\
& \displaystyle\sum_{j=1}^n y^j_{il}-nz_{il} \leq \frac{n}{2} & i\in \{2, \ldots, m\}, \; l\in \{1, \ldots, m\} \label{eq:bec5}
\\
& \displaystyle\sum_{q=1}^m qz_{1q}+\left(n-l\right)z_{il} \leq n-1 & i\in \{2, \ldots, m\}, \; l\in \{1, \ldots, m\} \label{eq:bec6}
\\
& \displaystyle\sum_{q=1}^m z_{1q} \geq 1 \label{eq:bec7}
\\
& x_i \in \left\{0,1\right\} & i \in \{1, \ldots, m\} \label{eq:bec8}
\\
& y^j_{il} \in \left\{0,1\right\} & j \in \{1, \ldots, n\}, \; i,l \in \{1, \ldots, m\} \label{eq:bec9}
\\
& z_{il} \in \left\{0,1\right\} & i,l \in \{1, \ldots, m\} \label{eq:bec10}
\end{flalign}
\vspace{-5mm}
\caption{Binary Integer Program for Bucklin Elections (Deleting Candidates)}
\label{BIPBUCKLINXX}
\end{figure}

Here (\ref{eq:bec0}) defines the objective function as the maximum 
number of candidates admitted to take 
part in the election. Each of constraints (\ref{eq:bec1}) permits the 
candidate $c_i$ to get a 
personal score of value $l$ iff the number candidates preferred by the 
voter $v_j$ over $c_i$ is 
strictly less than $l$. The form of (\ref{eq:bec1}) is the outcome of
\[
\displaystyle\sum_{k=1}^m A^j_{ki} x_k + y^j_{il} -m\left(1-y^j_{il}\right)\leq l,
\]
where the first term defines for voter $v_j$ the number of candidates 
which dominate $c_i$, while the remaining
part introduces the trigger variable $y^j_{il}$. Each of constraints (\ref{eq:bec2}) implies 
that candidate $c_i$ gets the personal score $l$ from voter $v_j$ when 
ranked among the top most $l$ 
candidates and it participates in the election. Indeed, the form of (\ref{eq:bec2}) is 
equivalent to
\[
\displaystyle\sum_{k=1}^m A^j_{ki} x_k + my^j_{il} +m\left(1-x_i\right) \geq l.
\]
Subsequently, for all $i \in \{2,\ldots,m\}$ constraints (\ref{eq:bec3}) 
restrict each candidate 
$c_i$ to get any personal scores when it is not allowed in the election. 
For all $i,l \in \{1,\ldots,m\}$ the set of constraints 
(\ref{eq:bec4}) allows candidate $c_i$ to earn the score $l$ iff it obtains 
strictly more than a half of all votes. In fact, the form of (\ref{eq:bec4}) 
results from
\[
\frac{1}{2} + \frac{n}{2} - \displaystyle\sum_{j=1}^n y^j_{il} + nz_{il} \leq n,
\]
where the first constant ensures the strictness concerning the half of the 
total number of votes given to $c_i$, 
the second term defines the half of all votes, the third term calculates 
the number of voters ready to give 
the score $l$ to $c_i$, and finally the combination of the fourth term 
and the right hand side constant introduces 
the trigger variable $z_{il}$ to handle the corresponding situation. 
In its turn, constraints (\ref{eq:bec5}) 
force each indicator $z_{il}$ to be equal to 1 every time when $c_i$ earns 
votes of more than a half of voters, 
for all $i \in \{2,\ldots,m\}$ and all $l \in \{1,\ldots,m\}$. 
When $z_{il}$ reveals that the given threshold is reached, (\ref{eq:bec6}) 
requires the minimal score 
obtained by the first candidate $c_1$ to be strictly less than the 
scores obtained by any other candidates. 
The form of (\ref{eq:bec6}) is equivalent to
\begin{equation}
\displaystyle\sum_{q=1}^m qz_{1q} +1 \leq lz_{il} + n \left(1-z_{il}\right),\label{eq:becNote}
\end{equation}
where the left hand side defines the minimal score obtained by $c_1$ and 
ensures the strictness of the inequality 
of the voting rule, while the right hand side defines the score obtained by candidate $c_i$ and bounds the former.
Note that constraints (\ref{eq:bec5}) are imposed for all the candidates except $c_1$. 
Therefore, variable $z_{1l}$ may not take the value of 1 for multiple scores $l$ in (\ref{eq:bec5}), $l=1 \ldots m$, while variable $z_{il}$ for $i=2 \ldots m$ must do this.
Along with constraint (\ref{eq:bec7}), this forces only one of the variables $z_{1q}$ be equal to 1 in the sum in the left hand side part of (\ref{eq:becNote}). It is the case corresponding to one of the possible scores $q$ for $c_1$ (at least the smallest one) as the left hand side part is bounded by the right hand side part. Constraint (\ref{eq:bec7}) implies that $c_1$ must get at least one 
of the scores, thus at least one variable $z_{1q}$ must be set to 1. 
In fact, it guarantees that at least the indicator pointing to the least 
score obtained by $c_1$ will trigger. 
Finally, the sets of constraints (\ref{eq:bec8}), (\ref{eq:bec9}) and 
(\ref{eq:bec10}) declare $x_1,\ldots,x_m$, $y^1_{11},\ldots,y^n_{mm}$ 
and $z_{11},\ldots,z_{mm}$ as binary. 

There always exists a feasible solution for the (\BEC) program which is guaranteed 
by deleting all possible candidates but $c_1$.

\subsection{Transition from Constructive Control to Destructive Control}\label{SubSec46}

At the beginning of Section~\ref{Sec4} we have promised to sketch how our 
models proposed for problems of constructive control can be adopted for ones of destructive control. 
Here, we explain the possible transition by the example of range voting.
Recall the decisive constraints (\ref{eq:sbe1}) of the program (\SBE).
If we combine these constraints into the single formula
\begin{equation}
\left(\sum_{j=1}^n A_{1j}x_j - \sum_{j=1}^n A_{2j}x_j \geq 1\right) 
\wedge
\ldots
\wedge
\left(\sum_{j=1}^n A_{1j}x_j - \sum_{j=1}^n A_{mj}x_j \geq 1\right),
\label{SBEC}
\end{equation}
then the binary decision vector $x \in \{0,1\}^n$ of the program (\SBE)
represents a subset $V^*$ of the set of voters $V$ such that candidate $c_1$ wins the 
range election $(C,V^*,(s_v)_{v \in V^*})$ iff (\ref{SBEC}) 
holds.
As a consequence, $c_1$ is not the winner of $(C,V^*,(s_v)_{v \in V^*})$ iff
the negation of (\ref{SBEC}) holds, or, equivalently, iff
\begin{equation}
\left(\sum_{j=1}^n A_{1j}x_j - \sum_{j=1}^n A_{2j}x_j \leq 0\right) 
\vee
\ldots
\vee
\left(\sum_{j=1}^n A_{1j}x_j - \sum_{j=1}^n A_{mj}x_j \leq 0\right)
\label{SBED}
\end{equation}
is true.
Such constraints with disjunctions are frequently called $k$-fold 
\emph{alternative constraints}, where $k$ defines the least number 
of constraints of the set which must be satisfied. In fact, 
mathematical programs with alternative constraints are no longer linear.
But there is a standard technique to transform alternative constraints
into a set of equivalent linear constraints; for details, see Chapter 9 
of \cite{Bradley}.
In our case it uses $k=1$ and replaces (\ref{SBED}) by
\begin{flalign}
~~~ & \sum_{j=1}^n A_{1j}x_j - \sum_{j=1}^n A_{ij}x_j \leq M \left(1-y_{i}\right)
 & i \in \{2,\ldots,m\}   
 & \label{SBEDKON1}
\\
 & \sum_{i=2}^{m} y_i \geq 1
 & \label{SBEDKON2}
\\
& y_i \in \left\{0,1\right\}    & i \in \{2, \ldots, m\} \label{SBEDKON3}
\end{flalign}
where each of auxiliary binary variables $y_2,\ldots,y_m$ reflects the satisfaction of 
the corresponding inequality of (\ref{SBED}), and $M$ is a large constant. In fact,
$y_i=1$ when candidate $c_1$ obtains less or equal number of scores than candidate $c_i$. 
The constant $M$ ensures that for all $i \in \{2,\ldots,m\}$
the formula $\sum_{j=1}^n A_{1j}x_j - \sum_{j=1}^n A_{ij}x_j \leq M \left(1-y_{i}\right)$
holds if $y = 1$. Here $M$ is set as 
$M=n \cdot max_{i \in \{1,\ldots,m\}, j \in \{1,\ldots,n\}} A_{ij}$. 
The constraint (\ref{SBEDKON2}) 
implies that at least one inequality of (\ref{SBED}) holds, and therefore there 
exists at least one candidate with scores equal or greater than candidate 
$c_1$'s. In such a way, to obtain the model for the destructive control 
problem one needs to replace constraints (\ref{eq:sbe1}) of the (\SBE) model 
by (\ref{SBEDKON1}) and (\ref{SBEDKON2}), and add those of (\ref{SBEDKON3}).

To adopt other proposed models in order to tackle the destructive control problems, 
it is necessary to substitute constraints responsible for the winner's determination 
with corresponding $k$-fold constraints.

\section{Computational Investigation}\label{Sec5}

In this section, the performance of the integer programs proposed in 
Section~\ref{Sec4} is evaluated in terms of their solution quality and the
needed computation time. We have implemented the program code in the programming
language JAVA using the 
\textsc{Cplex} 12.6 library with default settings such as 1e-6 for the 
feasibility and optimality tolerances and 1e-5 for the integrality 
tolerance. The experiments have been carried out on a desktop 
PC with an Intel Core i7 processor with 2.0 GHz and 8 Gb RAM. 

In the following, the efficiency of the proposed integer programming election
models is evaluated via PrefLib, assembled by (\cite{PrefLib}). 
PrefLib is a centralized repository of real world preference data.
Currently this library holds over 3000 datasets in an easily available and
computer readable form. 
The data comes from various sources and locations.
They are based on real elections (e.g., from Aspern, Berkley, Dublin, 
Glasgow, and trade unions in the European Union), real competitions 
(like the women's 1998 skating world championships and cross country 
skiing from the 2006-2009 world championships), comparison of websearches 
(across Bing, Google, Yahoo and Ask), reviews of hotels and etc.
Despite the voters of each instance of the library are grouped according 
to the equality of their preference lists, in our experiments we treat 
them independently assigning each one a separate decision variable. Furthermore, we select the first candidate 
of the list of candidates provided by each instance as the target winner $c_1$.

\subsection{Range Voting}\label{SubSec51}

To evaluate our range election model, 
we adopt the instances of the ``Tied Order - Complete List'' benchmark 
suite of PrefLib (\cite{PrefLib}). 
Within this suite both the relation of equivalence and the strict order 
relation between the candidates may exist for each of the voters. 
For every instance of the suite we 
construct the election as Borda-like election\footnote{%
  Note, that the elections are not Borda elections since
  the instances allow ties.}, i.e., by 
assigning the score 
of value $m-1$ to the most 
preferred candidates of a voter, $m-2$ to ones at the second place, and so on, 
while the least score of 0 is given to the candidate at the last place $m$ in 
a linearly ordered preference list (without ties) only. 
The up-to-date version of the benchmark suite 
consists of 331 instances. 
The largest instance in terms of the number of voters 
contains 299\,664 voters and 5 candidates, while the largest instance in terms of the 
number of candidates has 2\,819 of them and 4 voters. 
Exactly 293 of the 331 instances have 
been solved either to optimality or to optimality respecting the optimality 
tolerance. 
For each of the other 38 instances the infeasibility of the solution concerning the 
first candidate as a winner has been detected. 
The specific reasons that result 
in infeasibility are discussed in Section~\ref{SubSec41}. 
The computation time per instance 
is at most 3.5 seconds, while the median over all instances is 0.016 seconds 
only. 
In fact, all the range election instances from the presented benchmark 
suite can be easily solved. 
The total computation time over the whole suite is 25 seconds.

\subsection{Preference-based Voting}\label{SubSec52}

To test the other proposed integer programming 
election models of Section~\ref{Sec4} we employ the instances of the ``Strict Orders - 
Complete List'' benchmark suite of PrefLib (\cite{PrefLib}). 
Within this suite only a strict order relation between each pair of candidates is 
given for every voter such that the candidates are linearly ordered by each 
single voter. 
The suite contains 627 instances in total. 
The largest instance in terms of the number of 
voters has 14\,081 of them and 3 candidates, while the largest instance in terms of 
the number of candidates has 242 of them and 5 voters. 
Table \ref{tab:res} 
contains the information concerning the computation times required by the 
\textsc{Cplex} solver 
to find the optimal solutions for the instances of the Condorcet, plurality, maximin 
and Bucklin election models. The whole set of test instances is partitioned 
into four classes according to the number $m$ of candidates. Thus, the first 
class contains instances whose number of candidates $m$ falls into the range 
from 1 to 9, while the second, third and fourth classes have the number of 
candidates in the ranges 10-99, 100-199, and greater than 200, respectively. The classes 
correspond to the last four columns of the table one-to-one. 
The first four rows of the 
table report the number of candidates $m$, the number of voters $n$, the 
median $n'$ of the number of voters, and the number of instances $c$ in each 
of the classes, respectively. 
Therefore, most of the instances are rather small in terms 
of the number of candidates. Specifically, 523 of them are contained in the first class. 
The remaining rows are grouped and present the minimum, 
median, average and maximum computation time over each of the classes for every 
election model we have presented.

\begin{table}[t]
\centering
\caption{Running times used by Condorcet, plurality, maximin and Bucklin election models}

\vspace{4mm}
\label{tab:res}
{\scriptsize
\begin{tabular}{r|l||r|r|r|r}
\hline\hline
          & $m$                & 1-9     & 10-99   & 100-199 & $\geq 200$ 		  \\
          & $n$                & 4-14081 & 4-5000  & 4-5     & 4-242            \\
          & $n'$               & 21      & 4       & 4       & 5                \\
          & $c$                & 523     & 77      & 21      & 6                \\
\hline
\multirow{4}{*}{\CE} 	     & min     & 0       & 0       & 0       & 0     \\
                           & median  & 0.015   & 0       & 0       & 0.015 \\
                           & average & 0.009   & 0.006   & 0.004   & 0.015 \\
                           & max     & 0.078   & 0.047   & 0.016   & 0.032 \\
\hline
\multirow{4}{*}{\PE} 			 & min     & 0       & 0.015   & 0.250   & 1.092 \\
                           & median  & 0.016   & 0.063   & 0.374   & 1.396 \\
                           & average & 0.111   & 4.310   & 0.368   & 1.547 \\
                           & max     & 7.441   & 324.919 & 0.530   & 2.152 \\
\hline
\multirow{4}{*}{\ME}       & min     & 0       & 0      & 0.109   & 2.246  \\
                           & median  & 0.016   & 0.031  & 0.203   & 16.622 \\
                           & average & 0.027   & 0.058  & 0.299   & 15.452 \\
                           & max     & 0.218   & 1.061  & 2.013   & 27.518 \\
\hline
\multirow{4}{*}{\BEV}      & min     & 0       & 0      & 0.905   & 11.762 \\
                           & median  & 0.031   & 0.125  & 1.763   & 48.298 \\
                           & average & 0.042   & 0.431  & 3.810   & 50.807 \\
                           & max     & 0.281   & 4.680  & 17.800  & 93.054 \\
\hline													
\multirow{4}{*}{\BEC} 	   & min     & 0       & 0.047    & 91.604  & 6950.672  \\
                           & median  & 0.032   & 2.450    & 238.494 & 12410.285 \\
                           & average & 2.297   & 86.243   & 282.090 & 14527.973 \\
                           & max     & 469.891 & 5135.537 & 978.516 & 30466.901 \\																									
\hline
\hline																
\end{tabular}
}
\end{table}

For the Condorcet election model of Section~\ref{SubSec42}, exactly 
605 out of 627 instances 
have been solved to optimality. 
For the remaining 22 instances infeasibility respecting the target win 
of the first candidate has been shown. It takes significantly less than 
a second to solve any of the instances, while the whole suite has been 
computed in 5.5 seconds. 

For the plurality election model of Section~\ref{SubSec43}, optimal solutions have been 
found for all instances of the suite. 
Here 613 instances, thus almost all of the 627 instances, require less 
than a second to be solved. 
However, one instance occurred that needs significantly more computation time 
comparing to the others. 
It has the largest $m n$ product, i.e., the value that strongly 
correlates with the number of constraints used by the model. 
Specifically, its $m n$ product 
is 50\,000 and it takes 324 seconds to find the optimal solution. The whole suite can 
be evaluated in 407 seconds.
 
For the maximin election model of Section~\ref{SubSec44}, exactly 605 out of 627 
instances have been solved to optimality, where 619 instances 
require less than a second of computation time. 
For the unsolved 22 instances infeasibility respecting the win of the 
first candidate has been proved. 
The maximum computation time over all instances results in 28 seconds.
In fact, only the instances of the last class corresponding to the
largest $m$ incur considerable computation time. 
The whole suite has been evaluated in 117 seconds.

For the Bucklin election model with deleting voters of Section~\ref{SubSec45}, 
again for 605 out of 627 instances optimal solutions have been obtained. 
Exactly 593 instances are computed rather fast; each within one second. 
Only the last two classes of instances with a larger value of $m$ are 
time-consuming. The whole suite for this model 
is solved in 430 sec.

The Bucklin election model with deleting candidates of Section~\ref{SubSec45} 
has shown to be considerably harder to solve. 
Despite of the fact that 433 of the 627 instances have been solved within a second, 
there are hard instances with a maximum computation time of around 8.5 hours. 
Specifically, all the instances with a number of candidates greater than 
100 are rather time-consuming. 
Even the instances with a small number of candidates but large 
number of voters require considerably more time than those of other election models.
This is mainly because of the increased number of auxiliary 
variables and constraints used for the problem representation.
For all instances the optimal solutions have been found in a total computation time of 
approximately 28 hours.

\section{Conclusion}\label{Sec6}

We have introduced various voting systems and types of control of elections 
and shown how hard control problems can be modeled as integer programs.
Using the solver \textsc{Cplex} and test suites from PrefLib 
we have demonstrated that the approach allows to treat also larger
instances successfully.
Our experiments show that a proven hardness result of a control type is not a 
secure protection for the fraudulent falsification of outcomes of elections using 
this type.

As a future work, we are interested in extensions of our approach to other 
kinds of voting systems and other kinds of illegal influences.
Examples for the first are \emph{fallback voting} and \emph{SP-AV}, examples 
for the latter are \emph{partition of the voters} and \emph{bribery}.
In respect thereof, of great value it may be to investigate related computing
methods like SAT-solving, constraint programming, and functional-logic 
programming.

\section{Acknowledgements} 
We want to thank the referees for their valuable suggestions which helped to 
improve the paper.
This research has been supported through ARC Discovery Project DP130104395.

\section{References}

\bibliographystyle{apalike}
\bibliography{references}

\end{document}